# "FIELD PROGRAMMABLE DSP ARRAYS" - A NOVEL RECONFIGURABLE ARCHITECTURE FOR EFFICIENT REALIZATION OF DIGITAL SIGNAL PROCESSING FUNCTIONS

Amitabha Sinha[1], Soumojit Acharyya[2],Suranjan Chakraborty[3] and Mitrava Sarkar[4]

[1,2,3,4]Department of Microelectronics & VLSI Technology,
School of Engineering & Technology, West Bengal University of Technology
Kolkata - 700064, West Bengal, India
[1]amitabha.sinha@wbut.ac.in, [2]acharyyasoumojit@gmail.com
[3]suranjan.wbut@gmail.com, [4]mitrava_17dec@yahoo.co.in

## *ABSTRACT*

*Digital Signal Processing functions are widely used in real time high speed applications. Those functions are generally implemented either on ASICs with inflexibility, or on FPGAs with bottlenecks of relatively smaller utilization factor or lower speed compared to ASIC. The proposed reconfigurable DSP processor is redolent to FPGA, but with basic fixed Common Modules (CMs) (like adders, subtractors, multipliers, scaling units, shifters) instead of CLBs. This paper introduces the development of a reconfigurable DSP processor that integrates different filter and transform functions. The switching between DSP functions is occurred by reconfiguring the interconnection between CMs. Validation of the proposed reconfigurable architecture has been achieved on Virtex5 FPGA. The architecture provides sufficient amount of flexibility, parallelism and scalability.*

## 1. INTRODUCTION

DSP functions [1], [2] are computationally intensive and exhibit spatial [3], [4] parallelism, temporal [5] parallelism or both. High speed applications like Software Defined Radio (SDR), satellite modems, HDTV etc. need very high performance that is not achievable with currently available DSP processors [6], [7]. Even though higher performance achievement, relatively lower cost and low power dissipation are the major advantages of ASICs, high degree of inflexibility restricts their usage for rapidly changed scenario in the current high end applications as mentioned above. On the other hand, mapping different DSP functions at run- time, dynamically reconfigurable FPGAs [8], [9] are becoming popular because of their flexibility and low risk factor. However, lower utilization factor due to wastage of area in SRAM based CLBs, higher cost and relatively lower performance due to complex interconnection and routing delay are the major bottlenecks of the FPGAs. Although, some of the FPGAs of virtex family offer DSP basic building blocks like Multiply and Accumulation (MAC) units but silicon utilization factor is not optimized for the LUT based architecture [10] of FPGA. The proposed FPDA architecture eliminates the drawbacks of FPGAs and ASICs. DSP functions are mainly of two types: 'Filter Functions' (FIR, IIR etc.) and 'Linear Transforms' (DFT, FFT, DCT, DWT etc.). Keeping these in views, this paper presents a novel reconfigurable DSP architecture which combines different DSP functions by interconnections among different CMs.

Section- II of the paper describes Distributed Arithmetic Principle which has been used to implement DSP functions (like FIR, IIR, DCT, DWT etc.) in the proposed architecture. Section- III of the paper describes different DSP functions and their implementation proposal in proposed architecture. Section- IV describes the detailed representation of "Reconfigurable Architecture". Section- V analyzes the performance with various simulations, implementation and comparison results and Section- VI concludes the paper.

## 2. Distributed Arithmetic Principle

DA is extensively used in computing the sum of products. Distributed arithmetic is an efficient method for computing the inner product operation. Mathematical derivation of distributed arithmetic is extremely simple; a mix of Boolean and ordinary algebra. The variable Y holds the result of an inner product operation between the data vector x and the coefficient vector a. In distributed arithmetic representation the inner product operation is given as follows:

$$Y = \sum_{j=1}^{B-1} \left[ \sum_{i=1}^{N} x_{ij}\, a_i \right] 2^{-j} + \sum_{i=1}^{N} (-x_{i0})\, a_i \qquad (1)$$

Where the input data words $x_i$ have been represented by the 2's complement number presentation in order to bound number growth under multiplication. The variable $x_{ij}$ is the jth bit of the $x_i$ word which is Boolean, B is the number of bits of each input data word and $x_{i0}$ is the sign bit. Distributed arithmetic is based on the observation that the function can only take 2N different values that can be pre-computed offline and stored in a look-up table. Bit j of each data $x_{ij}$ is then used to address this look- up table. This equation clearly shows that the only three different operations required for calculating the inner product. First, a look-up to obtain the value of the function, then addition or subtraction, and finally a division by two that can be realized by a shift.

DA is actually a bit serial computation process, slowness arrives in this approach. The memory size also is increased exponentially in case of serial DA. This problem has been eliminated by partitioning the input words. Those inputs enter in a parallel fashion and increase the speed. Parallel DA [11], [12], [13] architecture replaces multiplier block by adder tree.

## 3. DSP Functions And Proposed Implementation

### 3.1. Finite Impulse Response Filter

An FIR with constant coefficients is an LTI digital filter. The output of an FIR of length L, to an input series x[n] is finite version of convolution sum:

$$y[n] = \sum_{k=0}^{L-1} x[k]\, c[n-k] \qquad (2)$$

Where $c[0] \neq 0$ through $c[L-1] \neq 0$.

In Figure 1 and Figure 2, 16 tap FIR filter has been implemented using Parallel DA. LUT contents for DA FIR are f(c[n-k], x[n]). DA implementation is mostly attractive for lower tap filter due to memory space limitations. LUT access time is a bottleneck for speed of higher tap filters. A LUT of $2^{16}$ locations is needed to implement 16 tap FIR using DA. As FIR filter is a linear filter, outputs of the lower tap filters are summed up to get the output of a higher tap filter. This paper proposes FIR architecture with 32 numbers of $2^4$ LUTs that cause decrease in memory locations and fast execution at the cost of excess LUTs, registers and adders. Parallel implementation of inherently serial DA has been performed in the proposed architecture where each bit of each input enters in parallel to the LUTs (2 LUTs for a coefficient). Memory size reduced and speed enhanced in the proposed scalable FIR architecture. The throughput of FIR filter is constructed using DA irrespective of the length of the filter. The basic building blocks, needed to implement FIR filter are LUTs, adders and registers.

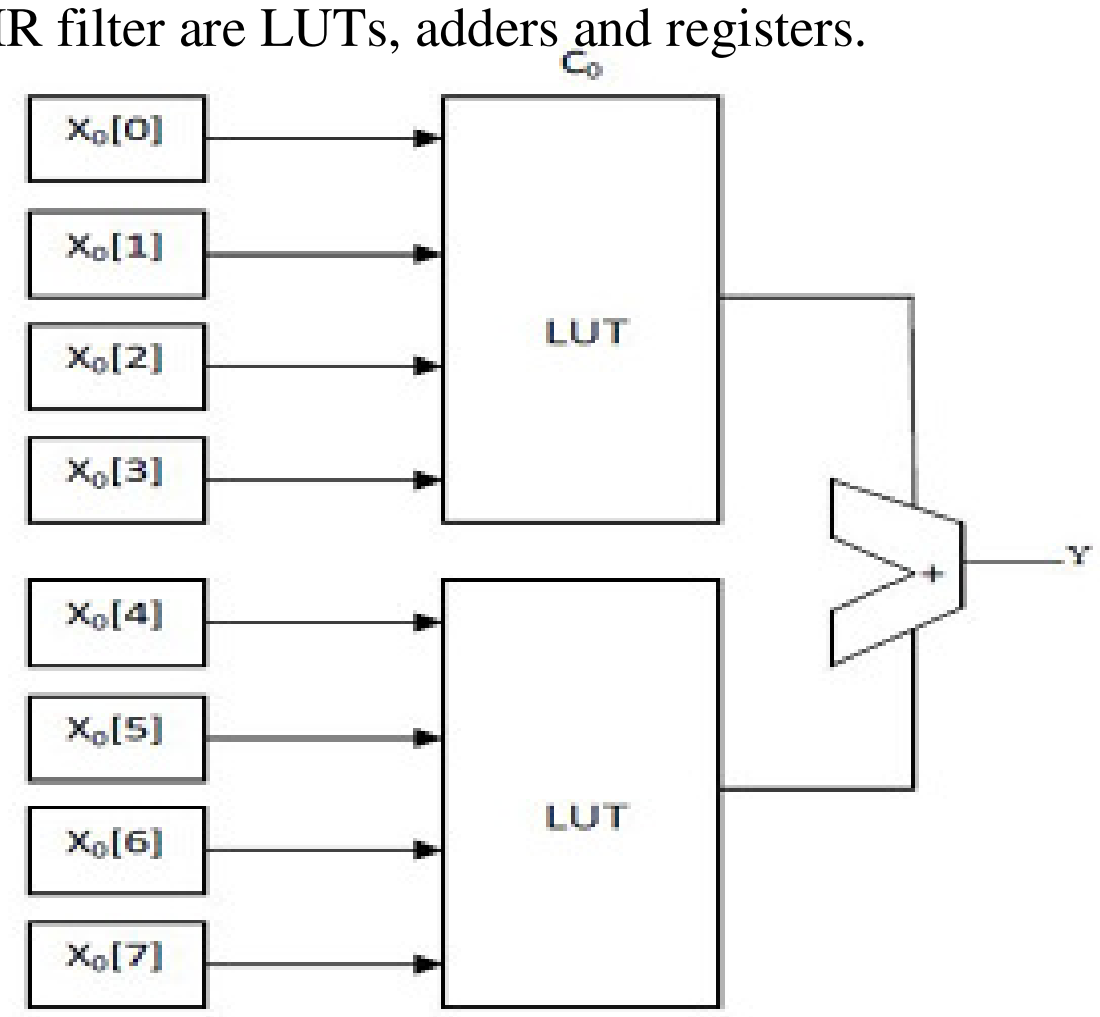

Figure 1. FIR unit for a coefficient

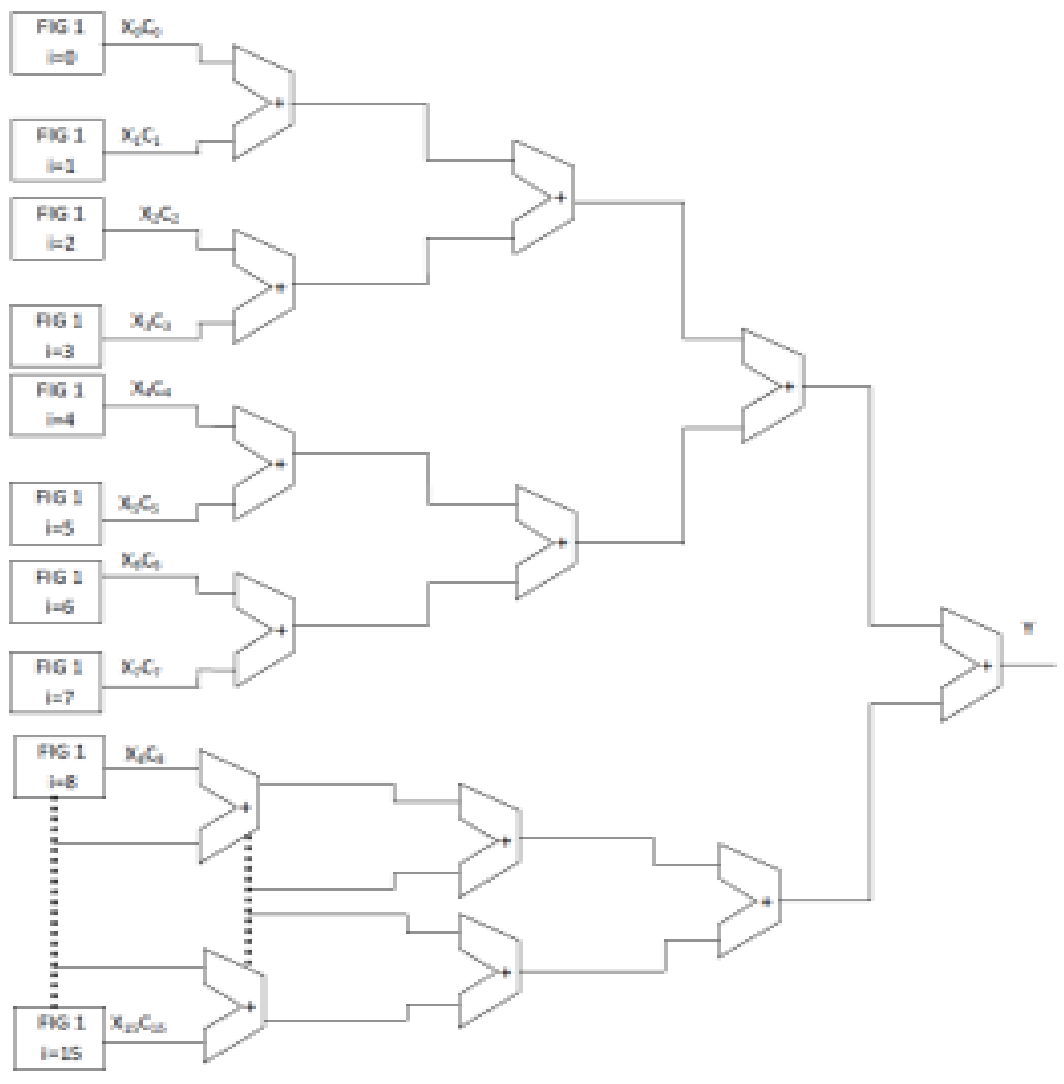

Figure 2. PDA FIR Filter Architecture

## 3.2. Infinite Impulse Response Filter

IIR filter is a recursive filter as it has feedback from output. The difference equation for such a system yield:

$$y[n] = \sum_{i=0}^{L-1} a[l]\, x[n-l] + \sum_{m=1}^{L-1} b[m]\, y[n-m] \qquad (3)$$

According to Figure 3, IIR is basically combination of two FIR filters and an adder. Implementation of IIR using Parallel DA has been done by forward filter and a feed backward filter. Feed backward filter is basically having the same input of forward filter with different LUTs. The basic building blocks needed to develop an IIR filter, are LUTs, adders and registers. For 3tap IIR filter:

$$y_2 = (a_0x_2 + a_1x_1 + a_2x_0) + (b_2y_1 + b_1y_0) \qquad (4)$$

Can be re written as:

$$y_2 = (a_0x_2 + a_1x_1 + a_2x_0) + \{x_0(b_2b_1a_0) + x_1(b_2a_0)\} \qquad (5)$$

From the above equations, it is observed that an IIR filter can be implemented using two FIR filters and an adder.

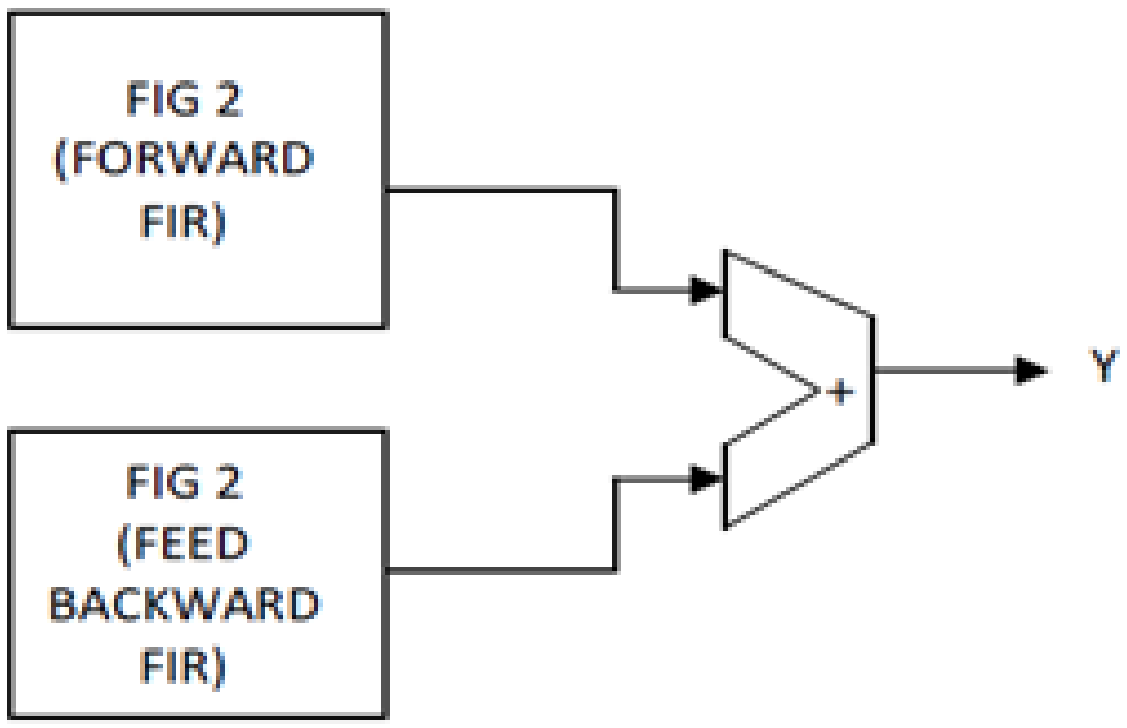

Figure 3. PDA IIR Architecture

## 3.3 Discrete Wavelet Transform

Discrete Wavelet Transform has been widely used in digital signal processing, image compression, data compression and communication domain in recent years. DWT is nothing but a system of filters. The two filters, "wavelet filter" and "scaling filter" are involved in DWT. The wavelet filter is a high pass filter whereas, scaling filter is a low pass filter. The coefficients of DWT are calculated recursively using Mallat's Pyramid Algorithm.

$$W_L(n, j) = \sum_m W_L(m, j-1)\, h_0(m - 2n) \qquad (6)$$

$$W_H(n, j) = \sum_m W_L(m, j-1)\, h_1(m - 2n) \qquad (7)$$

Where $W_L$ (n, j) and $W_H$ (n, j) are the $n^{th}$ scaling and wavelet coefficient at the $j^{th}$ stages, $h_0$ (n) and $h_1$ (n) are dilation coefficients corresponding to scaling and wavelet functions.

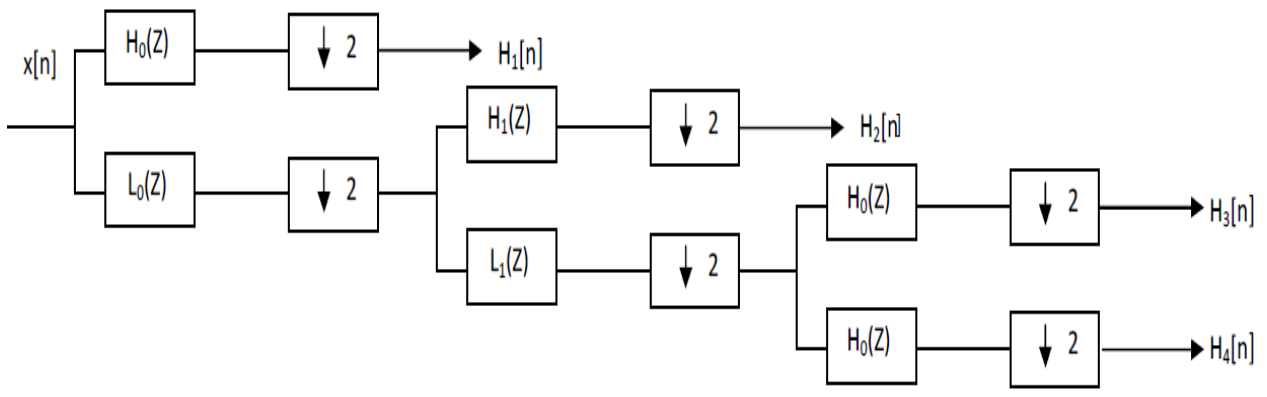

Figure 4. Discrete Wavelet Transform

The forward DWT has been implemented using Decimator block, which consists of a PDA FIR filter and a down sampling operator as shown in Figure 5. The PDA FIR has been implemented as FIR architecture described above in Figure 4. The FIR Daubechies 8- tap has been chosen for the implementation as shown in Table 1.

Table 1: Daubechies 8 Tap Filter Coefficients

| $H_0$ | L0 |
|---|---|
| -0.0106 | 0.2304 |
| -0.0329 | 0.7148 |
| 0.0308 | 0.6309 |
| 0.1870 | -0.0280 |
| -0.0280 | -0.1870 |
| -0.6309 | 0.0308 |
| 0.7148 | 0.0329 |
| -0.2304 | -0.0106 |

The FIR input has been driven by the clock i.e. synchronized with the clock input of the 1bit counter. The output port of FIR is connected to the input of parallel load register. Receive and obstruction of the inputs to the register, depend upon the state of the counter. The input enters decimator at the rate of 1sample/ clock while filtered output comes out at the rate of 1sample/ 2 clocks.

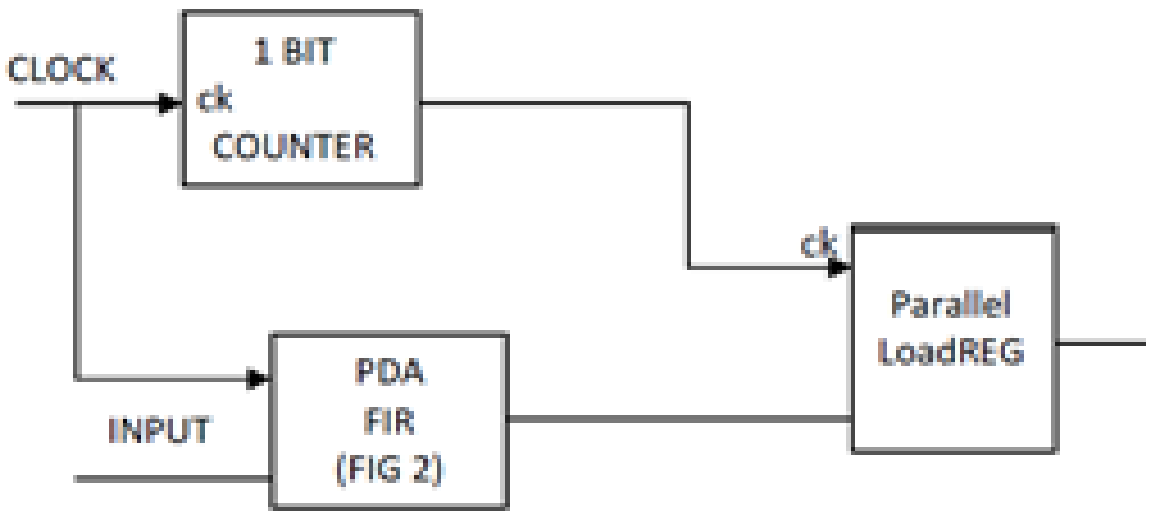

Figure 5. Implementation of Decimator

## 3.4. Fast Fourier Transform

Discrete Fourier Transform has played an important role in DSP. It is basically a discrete transform for Fourier analysis of the signal. The formulation of DFT for an input signal:

N - 1

$$X[k] = \sum_{n=0} x[n]\, e^{-(j2\pi kn/N)} \qquad (8)$$

FFT is basically a computation process of Discrete Fourier transform with multi-dimensional index mapping, suitable for real time application. The proposed FFT architecture has been implemented with decimation in frequency Cooley-Tukey algorithm [14], [15]. The efficient complex multiplier has been implemented for complex multiplication of butterfly in Figure 6.

$$R + jI = (a + ib)\,(\cos\theta + i\sin\theta) \qquad (9)$$

Final product of the complex multiplication:

$$R = (\cos\theta - \sin\theta)\, b + \cos\theta\,(a - b) \qquad (10)$$

$$I = (\cos\theta + \sin\theta)\, a - \cos\theta\,(a - b) \qquad (11)$$

Instead of cosine and sine table to compute complex multiplication, the implementation can be accomplished with three multipliers, one adder and two subtractors at the cost of one additional table, as shown in Table 2. The Butterfly has been implemented using proposed efficient complex multiplier.

Table 2 : Basic Blocks of Single Butterfly

| **Basic blocks of each butterfly unit** | **Number** |
|---|---|
| Adder | 1+1=2 |
| Subtractor | 1+2=3 |
| Multiplier | 3 |

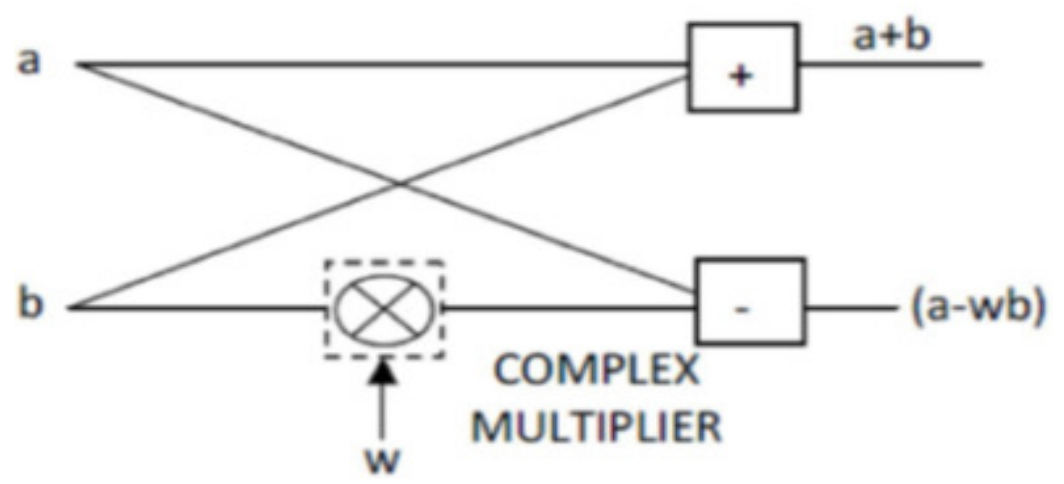

Figure 6. Butterfly Architecture

16 point proposed scalable FFT architecture has been implemented using eight butterfly units, sixteen registers, sixteen 4:1 multiplexers and fourteen 2:1 multiplexers in Figure 7.

The parallelism of the proposed architecture has been achieved by performing each stage with 8 butterfly units [16] that cause increase in speed. Output of stage n is input of stage (n+1). Output of butterfly unit fed back to the input. Multiplexer's select lines s0, s1 determine the stages while s2 incurs scalability to the proposed architecture.

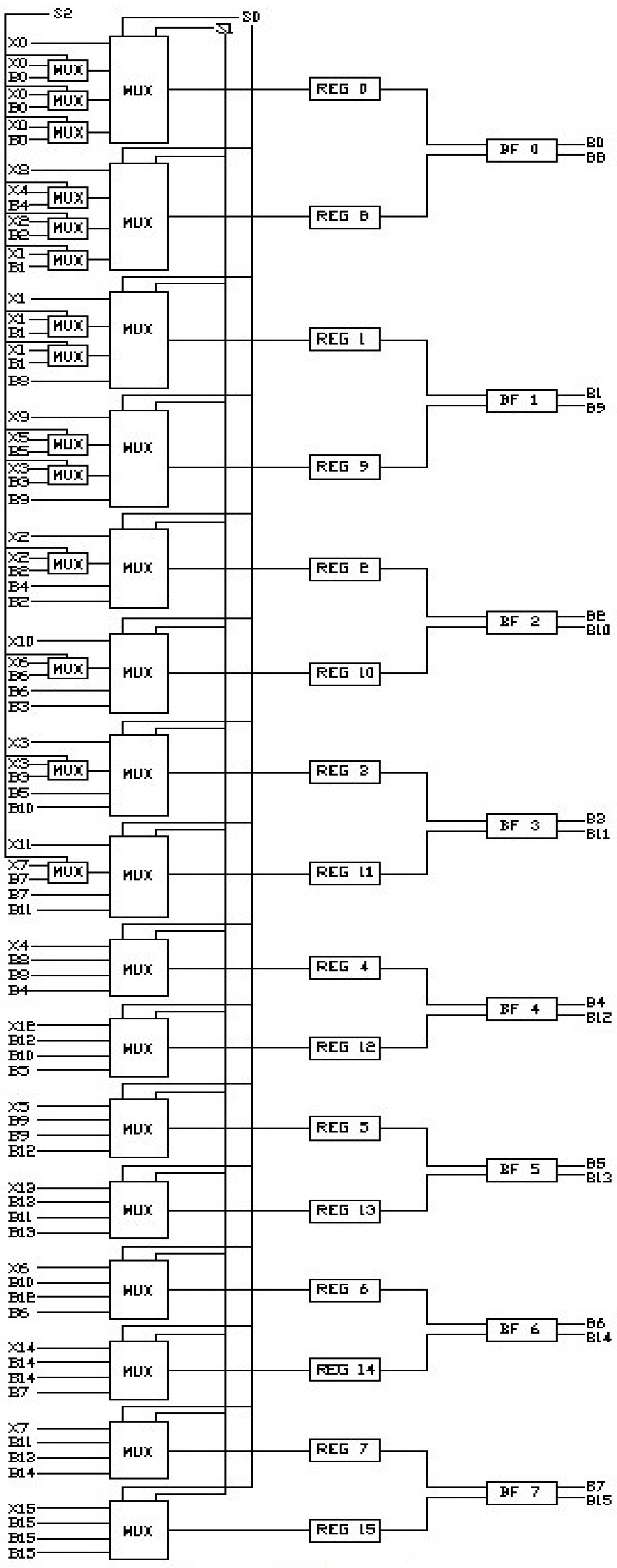

Figure 7. Scalable Proposed FFT Architecture

## 3.5. Discrete Cosine Transform

Discrete Cosine Transform is a Fourier related transform, dealing with real numbers only. 2D DCT is one of the efficient functions, used for different compression technique.

N point 1D DCT is defined by:

$$Y[K] = 2/N\,\{C_k \sum_{n=0}^{N-1} y(n) \cos[(2n+1)k\pi/2N]\} \qquad (12)$$

Where, $k = 0,1 \ldots\ldots N-1$

$C_k = \frac{1}{\sqrt{2}}$ When, k=0 else $C_k = 1$

The formula of 2D DCT can be computed by row-column decomposition of two 1D DCTs. 1D DCT blocks along row and column implement 2D DCT. In the proposed architecture of 1D Fast Discrete Cosine Transform, it has been implemented by Distributed Arithmetic [17], [18] in Figure 9. DCT constant coefficient for N = 16 can be represented as:

$$A = \cos\pi/4,\ B = \cos\pi/8,\ C = \cos 3\pi/8,\ D = \cos\pi/16,$$
$$E = \cos 3\pi/16,\ F = \cos 5\pi/16,\ G = \cos 7\pi/16,\ H = \cos\pi/32,$$
$$I = \cos 3\pi/32,\ J = \cos 5\pi/32,\ K = \cos 7\pi/32,\ L = \cos 9\pi/32,$$
$$M = \cos 11\pi/32,\ N = \cos 13\pi/32,\ O = \cos 15\pi/32$$

The matrix has been decomposed into even and odd subscript matrices. Even subscript matrix has been decomposed again into 4x4 matrices. Odd subscript matrix has been decomposed into a number of 4x4 matrices followed by adders, as shown in Figure 9.

Construction modules of Scaling Accumulator Unit are adder, substractor and shift register as in Figure 8.This unit acts like multiply and accumulate unit. Multiplication has been replaced with the Scaling Accumulator in proposed DA approach. LUT stores the multiplication value of DCT coefficient and each input bit. Output y is the multiplication result of input and coefficient.

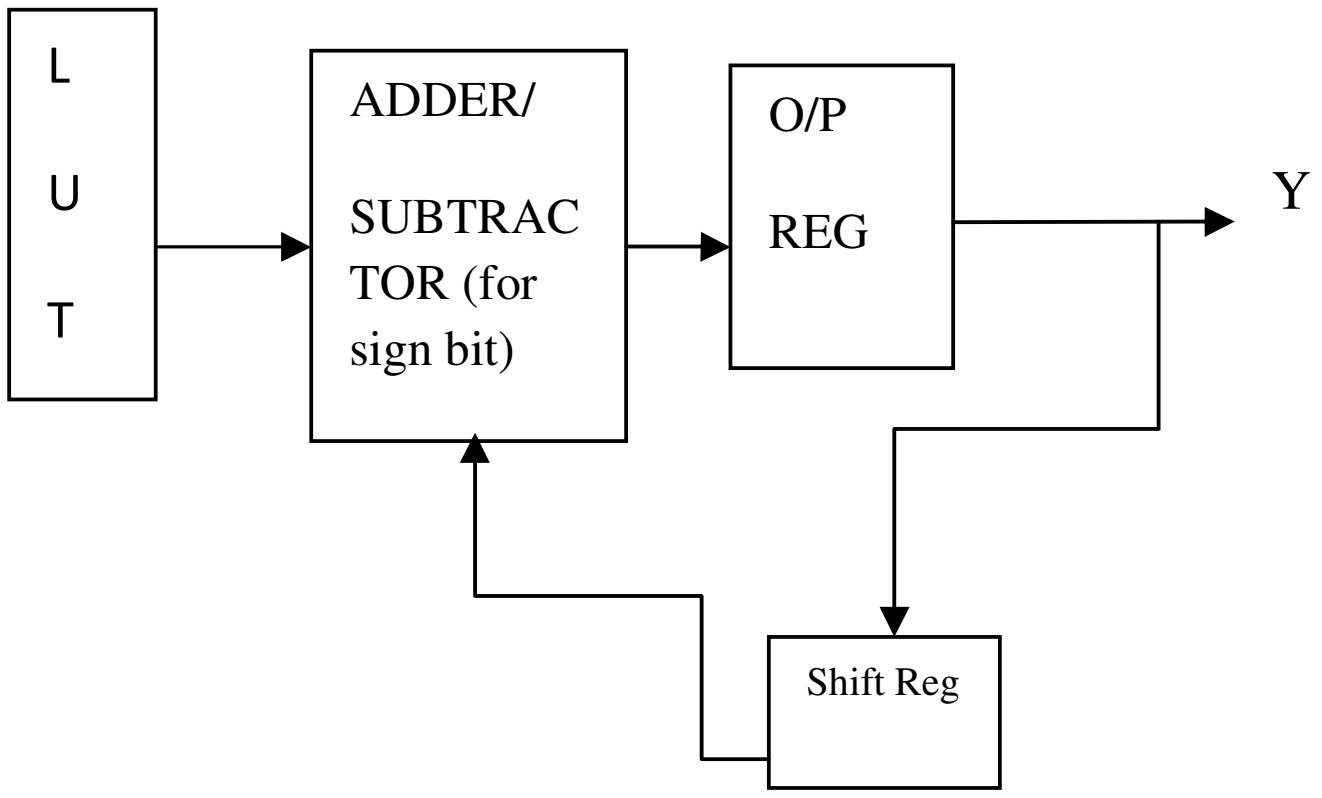

Figure 8. Scaling Accumulator Unit

Table 3 : Basic Blocks of a Single Accumulator

| **Blocks** | **Adders** | **Substractors** | **LUTs ($2^4$)** |
|---|---|---|---|
| Input Comb. Block | 12 | 12 | - |
| Decomposed Matrix | 32 | 24 | 24 |

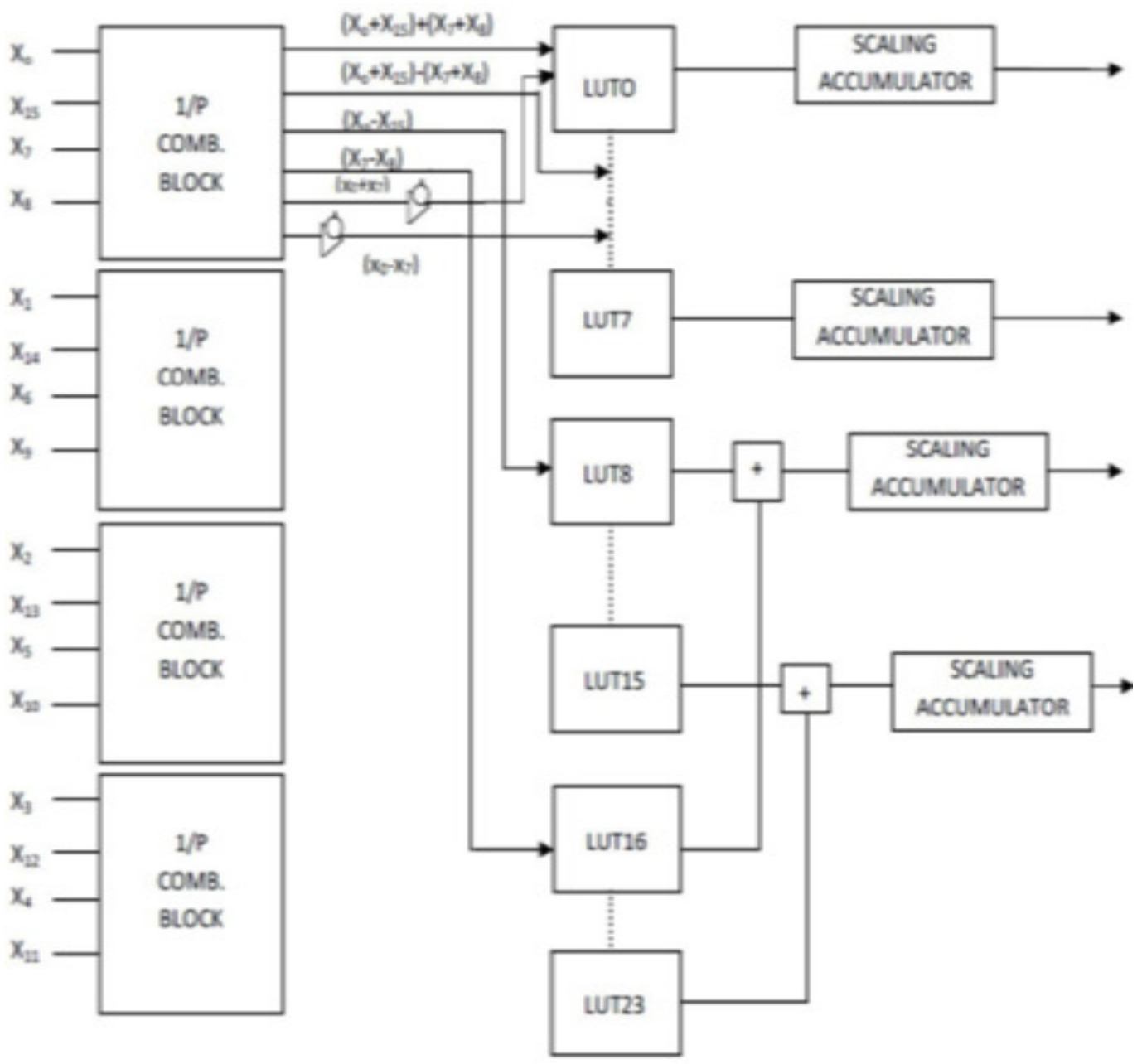

Figure 9. Scalable DCT architecture

$$\begin{bmatrix} Y0 \\ Y4 \\ Y8 \\ Y12 \end{bmatrix} = \begin{pmatrix} A & A & A & A \\ B & C & -C & -B \\ A & -A & -A & A \\ C & -B & B & -C \end{pmatrix} \begin{pmatrix} (x0+x15)+(x7+x8) \\ (x1+x14)+(x6+x9) \\ (x2+x13)+(x5+x10) \\ (x3+12)+(x4+x11) \end{pmatrix}$$

$$\begin{bmatrix} Y2 \\ Y6 \\ Y10 \\ Y14 \end{bmatrix} = \begin{pmatrix} D & E & F & G \\ E & -G & -D & -F \\ F & -G & D & E \\ G & -F & D & -E \end{pmatrix} \begin{pmatrix} (x0+x15)-(x7+x8) \\ (x1+x14)-(x6+x9) \\ (x2+x13)-(x5+x10) \\ (x3+12)-(x4+x11) \end{pmatrix}$$

$$\begin{bmatrix} Y1 \\ Y3 \\ Y5 \\ Y7 \\ Y9 \\ Y11 \\ Y13 \\ Y15 \end{bmatrix} = \begin{pmatrix} H & I & J & K \\ I & L & O & -M \\ J & O & -K & -I \\ K & -M & -I & O \\ L & -J & -N & H \\ M & -H & L & N \\ N & -K & H & -J \\ O & -N & M & -L \end{pmatrix} \begin{pmatrix} x0-x15 \\ x1-x14 \\ x2-x13 \\ x3-x12 \end{pmatrix} + \begin{pmatrix} L & M & N & O \\ -J & -H & -K & -N \\ -N & L & H & M \\ H & N & -J & -L \\ -O & -I & M & K \\ -I & K & O & -J \\ M & O & -L & I \\ K & -J & I & -H \end{pmatrix} \begin{pmatrix} x4-x11 \\ x5-x10 \\ x6-x9 \\ x7-x8 \end{pmatrix}$$

# 4. PROPOSED RECONFIGURABLE ARCHITECTURE

The proposed reconfigurable architecture consists of CMs like adders, substractors, multipliers, scaling units, registers etc. for 'Linear Transforms' (DWT, FFT and DCT) and 'Filter functions' (FIR and IIR) in Figure 10. Interconnection Matrix basically connects these CMs for a specific configuration. Different control signals define different configuration modes of the architecture, as shown in Table 4. A separate Decoder block has been introduced for generation of the control signal and selection of configuration mode. One configuration can be made at a time.

Table 4 : Control Signals for Different Configuration Modes

| **FUN.** \ **Control** | **C1** | **C2** | **C3** | **C4** | **C5** |
|---|---|---|---|---|---|
| FIR | 1 | 0 | 0 | 0 | 0 |
| IIR | 0 | 1 | 0 | 0 | 0 |
| DCT | 0 | 0 | 1 | 0 | 0 |
| FFT | 0 | 0 | 0 | 1 | 0 |
| DWT | 0 | 0 | 0 | 0 | 1 |

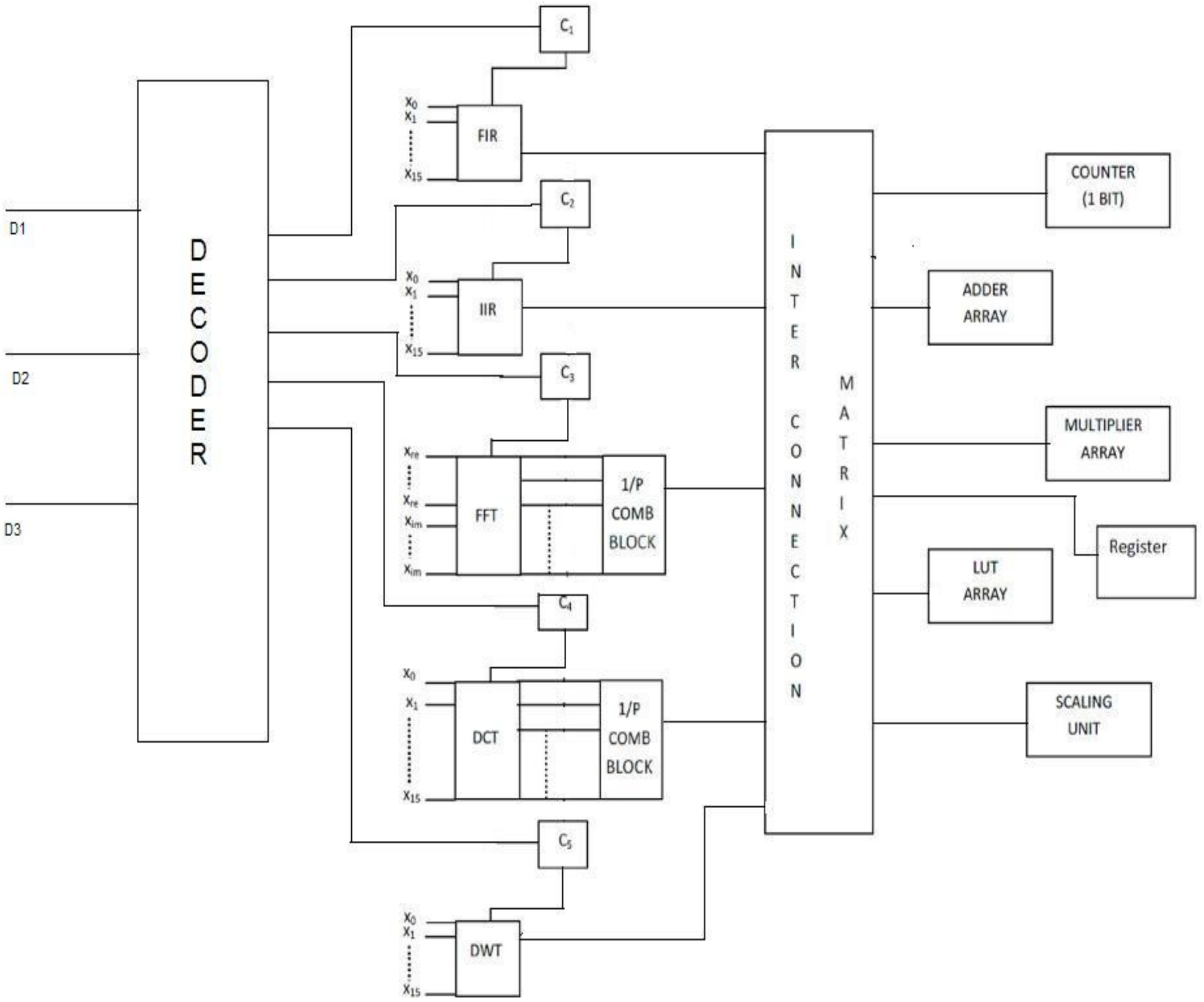

Figure 10. Proposed Reconfigurable Architecture

## 5. RESULTS AND ANALYSIS

The reconfigurable architecture has been validated on Virtex-5 FPGA. The synthesis report has been discussed below.

### 5.1 Final Reports

Selected Device: 5vlx330tff1738-2
Number of Slice Registers: 7476 out of 207360 3%
Number of Slice LUTs: 4686 out of 207360 2%
Number used as Logic: 4686 out of 207360 2%
Number of IOs: 886

Slice Logic Distribution:
Number of Bit Slices used: 8829
Number with an unused Flip Flop: 1353 out of 8829 15%
Number with an unused LUT: 4143 out of 8829 46%
Number of fully used Bit Slices: 3333 out of 8829 37%

Timing Summary:-
Speed Grade: -2
Minimum period: 4.937ns
Maximum Frequency: 202.558MHz
Minimum input arrival time before clock: 4.222ns
Maximum output required time after clock: 2.799ns

Specific Feature Utilization:
Number of BUFG/BUFGCTRLs: 3out of 32, 9%
Number of DSP48Es: 24 out of 192, 12%
Requirements: 7476 slices, 4686 LUTs, 886 IOB.

### 5.2 Analysis

Table 5 No. of Basic Blocks for DSP Functions

| Func | Counter | Ad der | LUT | sub | Regist er | MUX | Multipli er |
|---|---|---|---|---|---|---|---|
| FIR | - | 31 | 32 | - | 16 | 1(2:1) | - |
| IIR | - | 62 | 62 | - | 31 | 1(2:1) | - |
| DWT (DECIMATOR) | 1(1BIT) | 8 | 8 | - | 9 | - | - |
| FFT | - | 16 | - | 24 | 48 | 16(4:1)<br>14(2:1) | 24 |
| DCT | - | 44 | 24 | 36 | 32 | 8(2:1) | - |

The Reconfigurable Architecture has been implemented using only one 1bit counter, 62 adders, 94 LUTs, 36 subtractors, 24 multipliers as shown in Table 5. Reconfigurable architecture has been implemented using 4686 slice LUTs out of 207360 slice LUTs of FPGA whereas, FPDA is the alternative architecture to implement this combined architecture with only 94 LUTs. This Proposed DSP dedicated Reconfigurable Architecture combines different DSP functions here with optimized silicon area compare to FPGA.

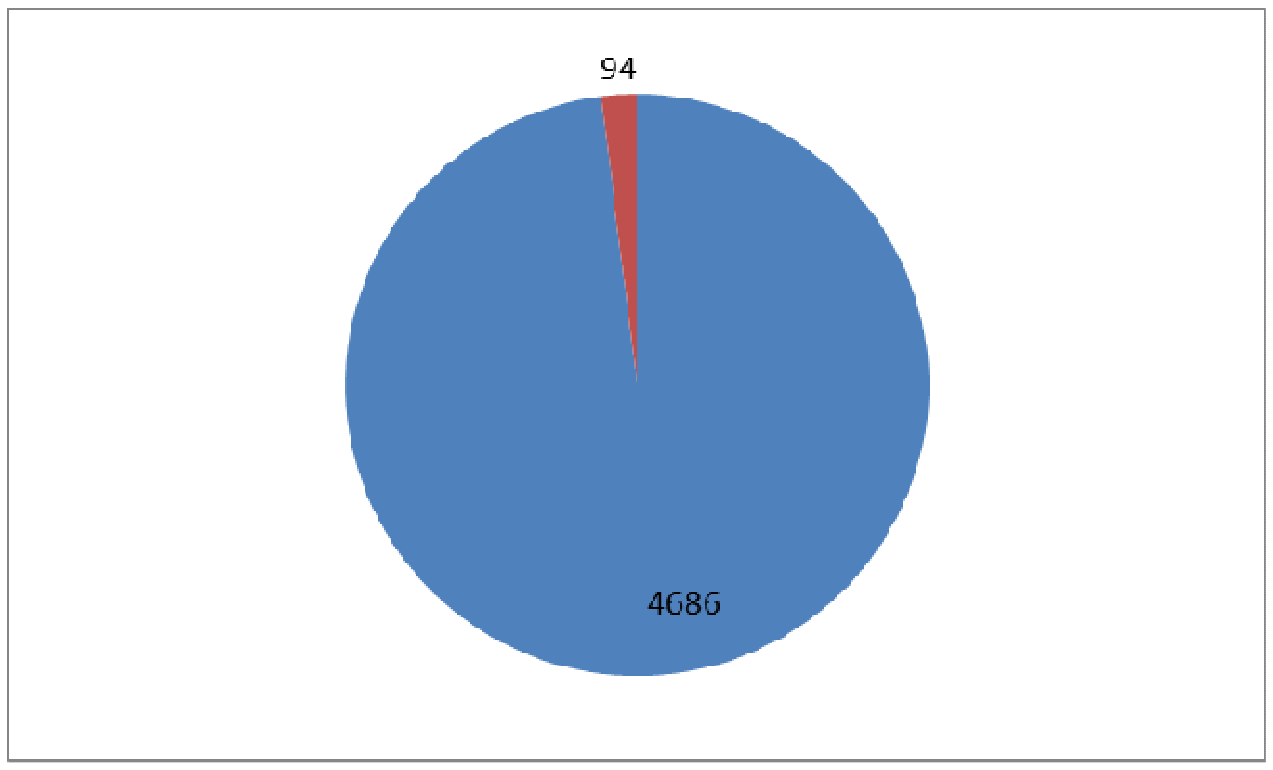

Figure 11. Comparison of Minimum no. of LUTs between FPGA and FPDA

In Figure 11, Blue indicates minimum number of LUTs which are needed to implement the combined architecture in FPGA. In contrast, Brown indicates the minimum number of LUTs that are needed to implement the same architecture in FPDA.

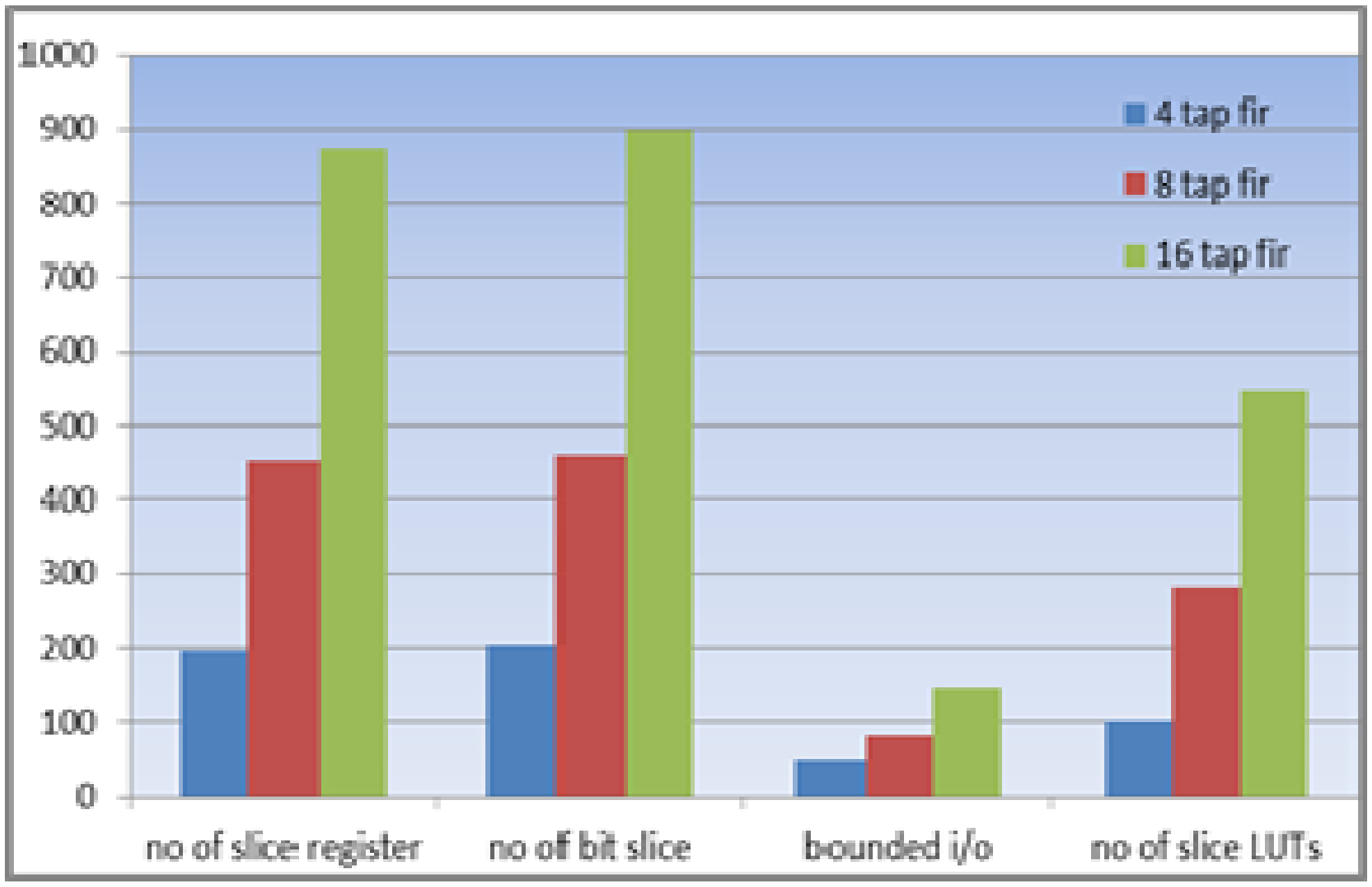

Figure 12. Comparison of device utilization factor for different tap FIR

Figure 12 shows device utilization factor for 4tap, 8tap and 16tap FIR filter for the architecture. Device utilization increases proportionally with number of taps.

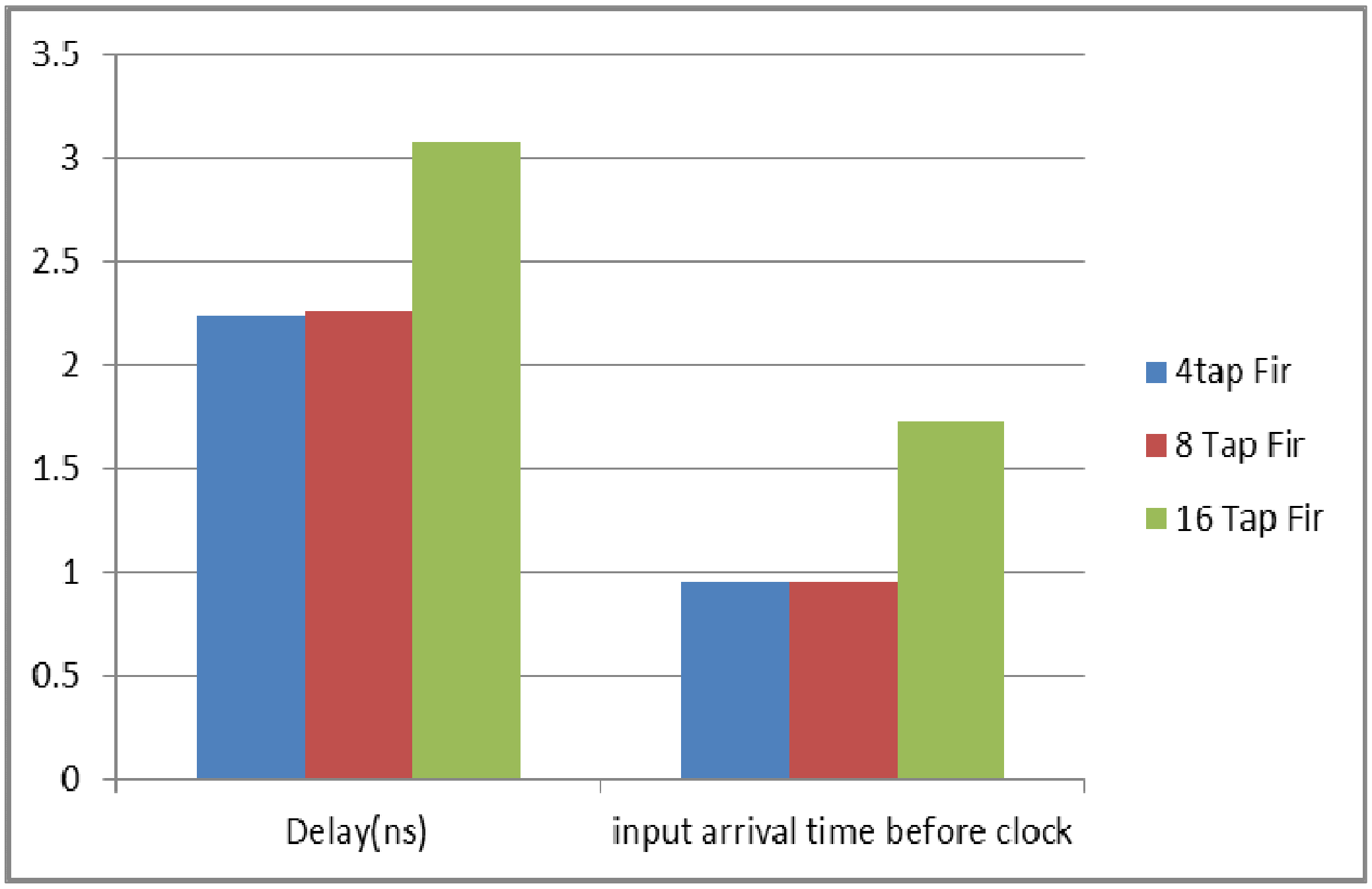

Figure 13. Comparison of timing for different tap FIR

Figure 13 shows comparison of timing for different taps of FIR filter. “Maximum output required time after clock” doesn’t depend on the tap number.

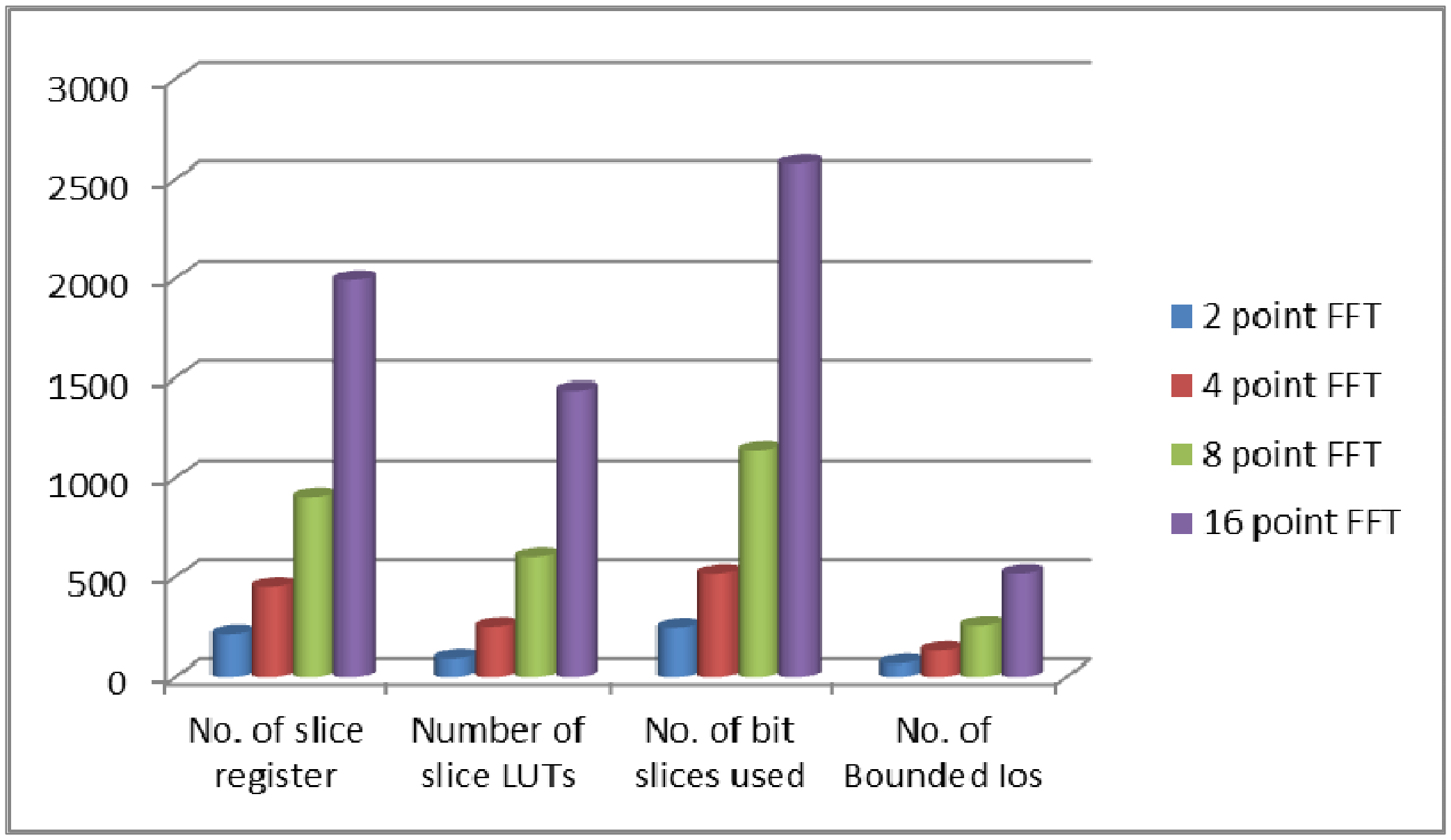

Figure 14. Comparison of device utilization factor for different point FFT

Figure 14 bar graph shows device utilization factor for different point FFT. Device utilization is proportional with the number of point FFT.

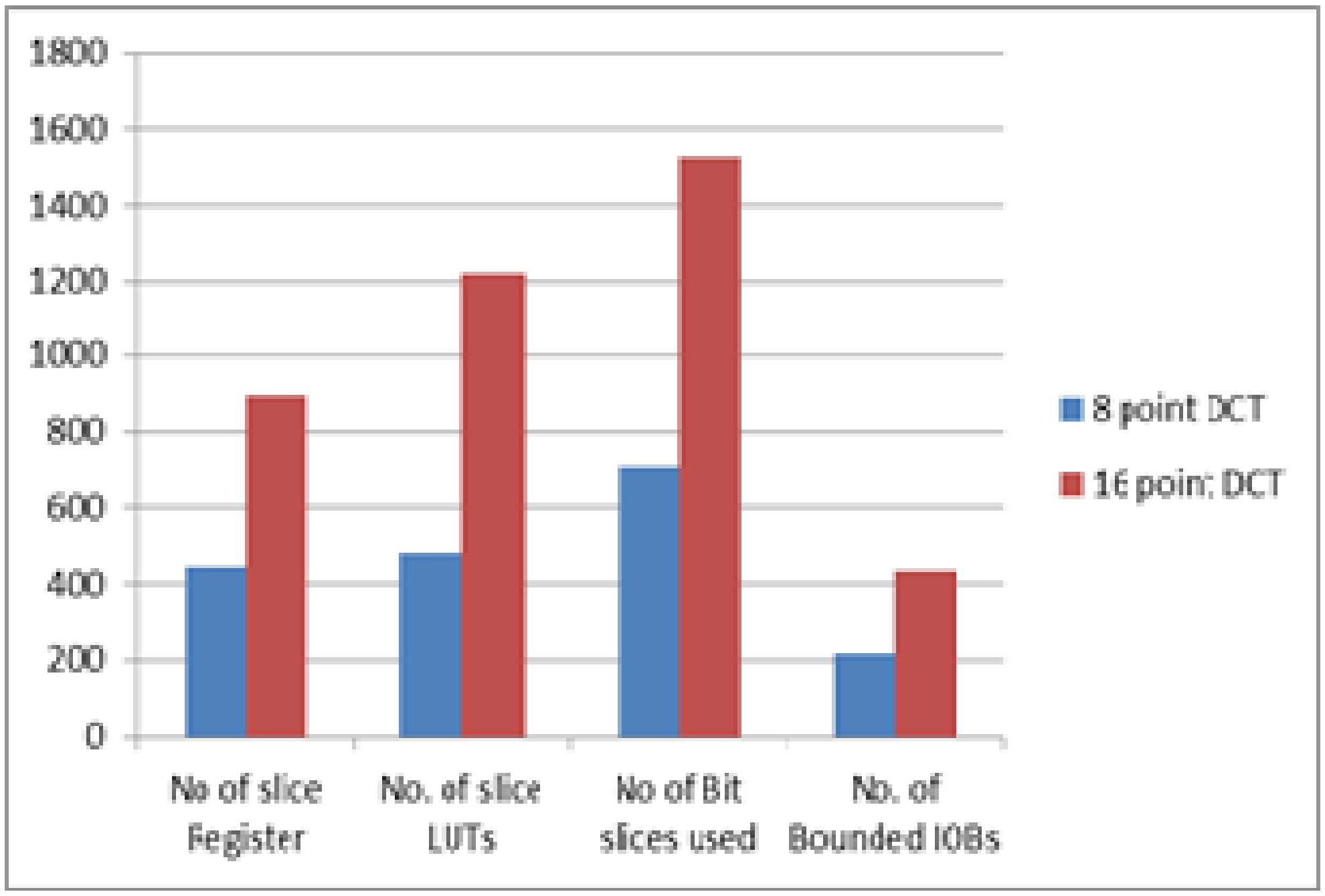

Figure 15. Comparison of device utilization factor for different point DCT

Similarly, Figure 15 shows comparison of device utilization factor for different point DCT.

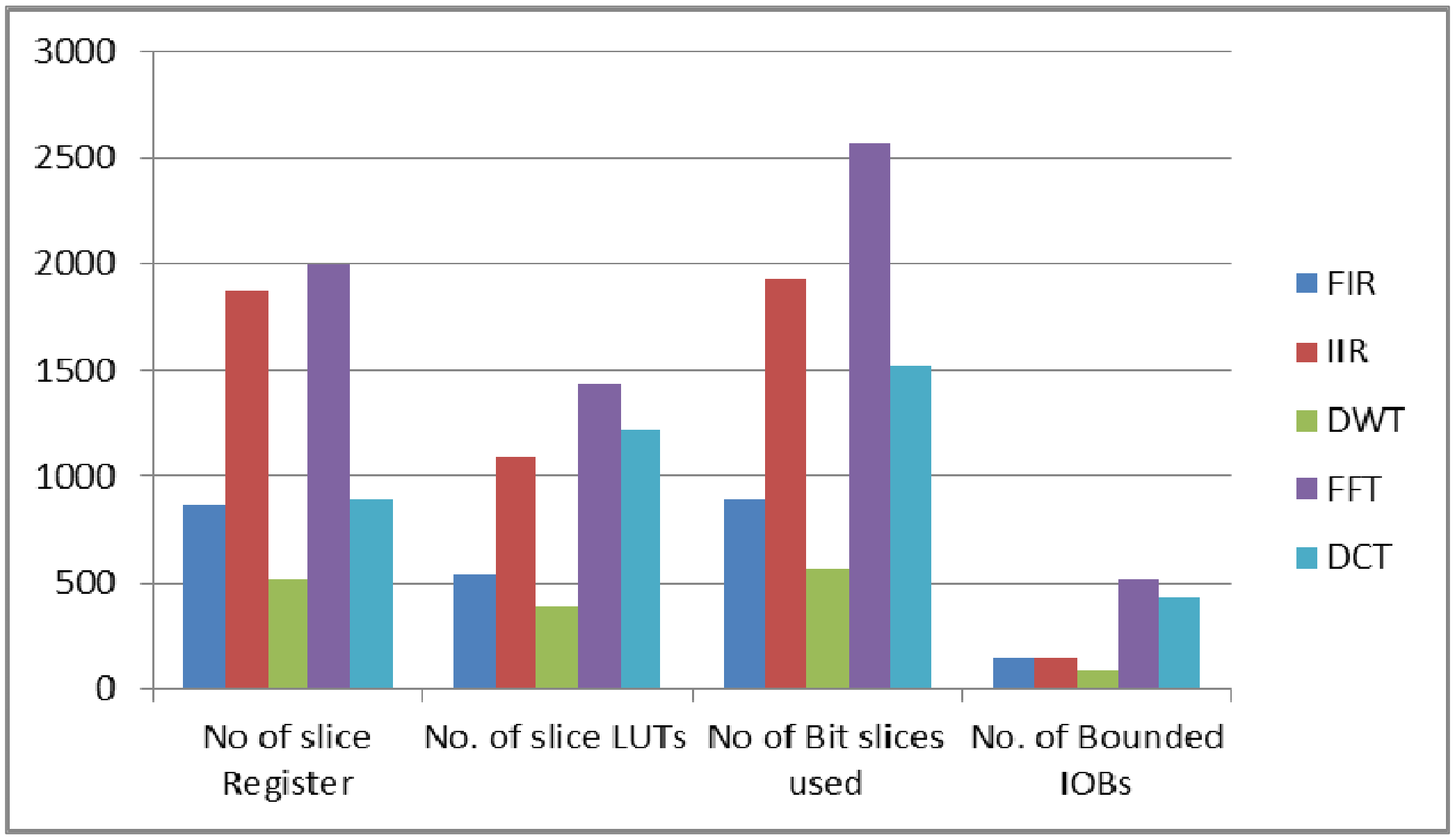

Figure 16. Comparison of device utilization factor

Figure 16 shows the comparison of device utilization factors of 'Combined Architecture' on virtex-5 FPGA with utilization of only 3% of slice register. Only 37% fully used bit slice Out of 8829. Number with an unused Flip Flop in bit slices is 15%. 46% of the bit slices with at least an unused LUT. It is clear from the above analysis, that implementation of DSP functions on FPGA has a serious bottleneck of lower utilization factor. "FPDA" is an array of basic common modules (CMs). Interconnections among those modules configure the device for a specific DSP function. The proposed reconfigurable architecture "FPDA" has relatively better utilization factor compare to FPGA.

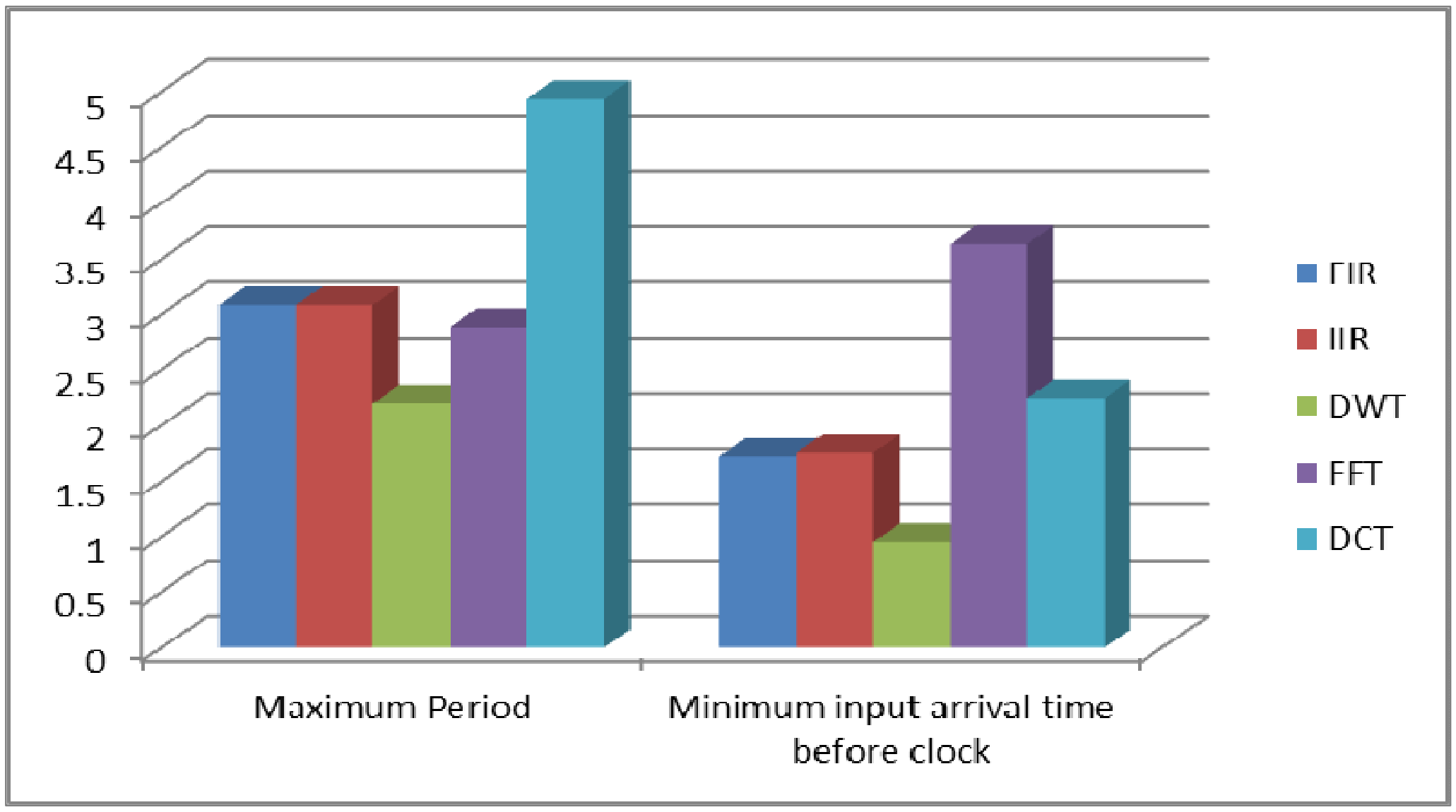

Figure 17. Timing comparison for combined architecture

Figure 17 shows the timing comparison between different DSP functions on the proposed architecture. Worst case timing from a flip-flop to other flip-flop through the logic within the FPGA is termed as "minimum period" i.e. 4.937ns for the proposed architecture when implemented on virtex-5 FPGA. "Minimum input arrival time before clock" is the worst case input data setup time requirement to clock pin has been reported 4.222ns. This minimum input arrival time before clock is maximum for configuration of DCT. The worst case data output delay after clock pin which is same in all cases, is termed as "Maximum output required time after clock" in the final report. No combinational data path from input to output. "FPDA" will offer high speed as configuration is basically interconnections among basic modules of DSP instead of complex interconnections in FPGA. There are so many advantages to realize DSP algorithms in the proposed FPDA architecture other than FPGA:

- different DSP functions can be made by changing the connectivity among the basic building blocks,
- placement & routing of basic building blocks in such a fashion that it should be optimum in delay than FPGA,
- architecture has a low design complexity,
- higher utilization factor than FPGA,
- high degree of parallelism and scalability.

## 6. CONCLUSIONS

The proposed "Reconfigurable Architecture" includes 'Filter Functions' and 'Linear Transforms'. The combined circuit is basically the union of all the basic building blocks mentioned above and they are required for implementing each of the functions. By interconnecting different building blocks in different fashions various DSP functions can be made. This process can be viewed as "Configuration". The architecture also offers scalability as new transforms with higher number inputs or higher tapped filter functions can also be implemented with those basic building blocks. The problems of inflexibility of ASICs, low utilization factor and low performance of FPGAs can be overcome with the proposed architecture as the major building blocks which are common to most of the DSP functions are implemented by direct hardware and not by LUT thereby optimizing the silicon utilization factor. Only one configuration can be made at a time which can be observed as a limitation of the architecture. But, minimization and maximum utilization of the hardware have been achieved at the cost of mentioned limitation. The future work can be proceed with

- the VLSI implementation of the proposed architecture,
- the implementation of different filter or linear transform functions in the proposed architecture and globalize the architecture for implementation of all DSP functions,
- implementation of high speed building blocks to achieve comparatively more faster architecture,
- time and hardware complexity analysis of the proposed hardware with other DSP functions and analysis the feasibility.

Employing Distributed Arithmetic approach for FIR, IIR, DWT and DCT functions and exploitation of the inherent parallelisms of the DSP functions, enhance the speed of the proposed architecture over the FPGAs substantially.

The Proposed architecture was validated on Xilinx virtex-5 FPGA on 5vlx330tff1738-2 using Xilinx ISE 9.1i.

## Authors

**Amitabha Sinha**

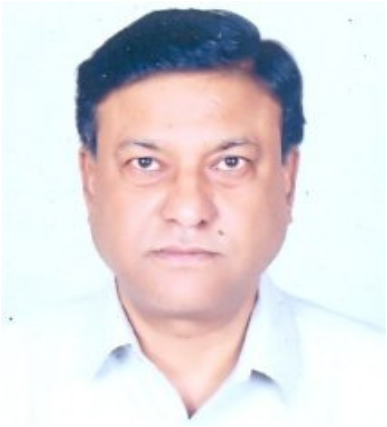

Prof.Amitabha Sinha has been working with Neotia Institute of Technology, an Institution affiliated to West Bengal University of Technology as a Principal. Prior to this assignment, he was with West Bengal University of Technology, Kolkata, India as a Professor and Director, School of Engineering & Technology since August 2005/2006.

He has more than 30 years of experience in industry, premier academic institutes, R&D centers and IT/Telecom organizations in India & abroad and his areas of research include Embedded System Design , VLSI design , Digital Signal Processing, Re-configurable Architecture using FPGAs, Software Defined Radio, Processor Architecture and System On-chip Design , Advanced Computer Architecture etc. He was Assistant Professor of the Dept. of Comp. Sc. & Engg., Bengal Engineering College ( Now Bengal Engineering & Science University), West Bengal, India during 1985-1991 and Associate Professor of the Dept. of Comp. Sc. & Engg , BITS Pilani, during 1997-1998. He was visiting faculty, Dept. of Computer Sc. & Engg., Oakland University, U.S.A. during 1994 and 1999. He was an expatriate faculty member of a premier Malaysian University Mara Institute of Technology during 1994-1995. Prof. Sinha also served R &D centres of different premier industries in the country at senior levels. With an Electronics & Comm. Engineering background, he holds a Ph.D degree in Computer Sc. & Engg. from Indian institute of Technology (IIT), Delhi which he had obtained in 1984.

**Soumojit Acharyya**

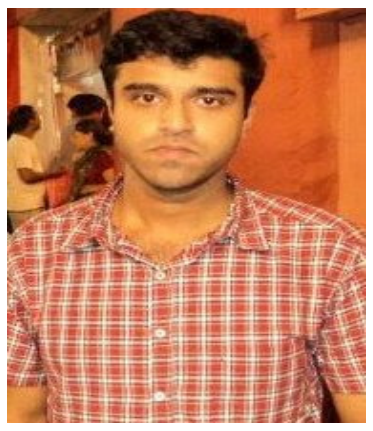

Mr. Soumojit Acharyya obtained his M.Tech degree in Microelectronics & VLSI Technology from West Bengal University of Technology and is working now at Wipro Technologies as a Project Engineer. His areas of expertise are Reconfigurable Architecture, VLSI Design and DSP (Digital Signal Processing).

**Suranjan Chakraborty**

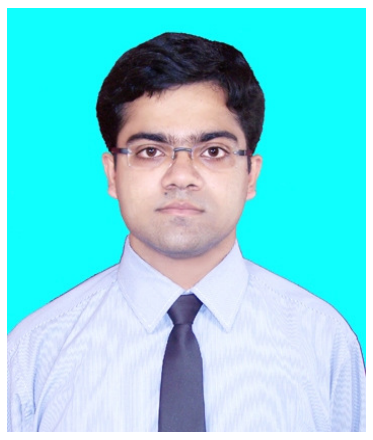

Mr. Suranjan Chakraborty obtained his M.Tech degree in Microelectronics & VLSI Technology from West Bengal University of Technology and is working now at Wipro Technologies as a Project Engineer. His areas of expertise are Application specific digital circuit design using FPGAs, Reconfigurable Architecture and DSP (Digital Signal Processing).

**Mitrava Sarkar**

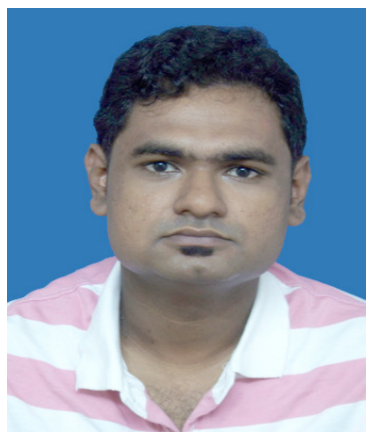

Mr. Mitrava Sarkar obtained his M.Tech degree in Microelectronics & VLSI Technology from West Bengal University of Technology and is working now at Infosys Ltd. as a System Engineer. His areas of expertise are Application specific digital circuit design using FPGAs and DSP (Digital Signal Processing).